\documentclass[11pt]{article}

\newcommand{\const}{{\nu}}

\def\tr{{\rm tr}}
\def\Tr{{\rm Tr}}
\def\I_M{{I_{\scriptscriptstyle M\times M}}}

\def\const{{\nu}}
\def\salpha{{\alpha}}

\begin{document}

\thispagestyle{empty}
\rightline{~~~KIAS-P02031}
\rightline{UOSTP-02102}
\rightline{{\tt hep-th/0204221}}

\vskip 2cm \centerline{ \Large\bf BPS equations in Six and Eight  Dimensions}

\vskip .2cm

\vskip 1.2cm

\centerline{ Dongsu Bak,$^a$  Kimyeong Lee,$^{b}$  and Jeong-Hyuck Park$^{b}$}
\vskip 10mm \centerline{ \it $^a$ Physics Department, University of
Seoul, Seoul 130-743, Korea} \vskip 3mm \centerline{\it $^b$ School of Physics, Korea Institute for Advanced Study} \centerline{ \it 207-43,
Cheongryangryi-Dong, Dongdaemun-Gu, Seoul 130-012, Korea } \vskip 1.2cm

\centerline{\tt  (dsbak@mach.uos.ac.kr,   klee@kias.re.kr, jhp@kias.re.kr)}

\vskip 1.2cm

\begin{quote}

We  consider the maximal supersymmetric pure Yang-Mills theories on six and eight dimensional space. We determine, in a systematic way, all the
possible fractions of supersymmetry preserved by the BPS states and present the corresponding  `self-dual' BPS equations. In six dimensions the
intrinsic one has 1/4 supersymmetry, while in eight dimensions, 1/16, 2/16, $\cdots$, 6/16. We apply our results to some explicit BPS configurations
of finite or infinite energy on commutative or noncommutative spaces.

\end{quote}

\newpage
\setcounter{footnote}{0}

\section{Introduction}

Recently there has been  considerable interest in understanding the possible supersymmetric states of D0 branes on D2, D4, D6 and D8 in IIA string
theories.  While the D0-D2 and D0-D4 systems are relatively well understood with or without the NS-NS  $B$ field background, the D0-D6 and D0-D8
cases remain veiled.

A pioneering work on higher  than four dimensional gauge theories was done by Corrigan, \textit{et al}. \cite{cdfn}. They investigated the higher
dimensional analogues of `self-duality', seeking linear relations amongst components of the field strength
\begin{equation}
F_{ab}+\frac{1}{2}T_{abcd}F_{cd}=0\,, \label{BPS}
\end{equation}
with constant four form tensor $T_{abcd}$. In four spatial dimensions it is essentially unique, i.e. $\pm\epsilon_{abcd}$. One immediate consequence
of the above equation is that the Yang-Mills field equations $D_{a}F_{ab}=0$ are automatically followed due to the Jacobi identity. The constant
four form tensor was introduced by hand in order to match the indices of the field strength. On the Euclidean space of dimension $D>4$, it inevitably
breaks the $\mbox{SO}(D)$ invariance and the resulting first order equations may be classified by the unbroken rotational symmetries
\cite{cdfn,ward}.

There have been much attention  to these higher dimensional `self-dual' equations. Especially notable ones include the
ADHM-like construction in $4k$ dimensions \cite{4kADHM} and the octonionic instantons in eight dimensions
\cite{fubini,harveystrominger}. The solutions constructed in this context have infinite energy. In fact, by using the
conservation of the energy momentum tensor, one can show that there exits no static finite energy solution
 in Yang-Mills theories on commutative space of  dimension $D>4$ \cite{jaffe}.

However, the above theorem does not apply to the noncommutative Yang-Mills  theories    essentially due to the scale symmetry breaking. In the
decoupling limit of string theory with a constant NS-NS $B$ field background, the field theories describing the worldvolume of D branes become
non-commutative \cite{non-comm}. Several localized  solutions with unbroken  supersymmetries have been found to have a finite energy or action in
the non-commutative gauge theories \cite{shift,witten,kraus}. Some exact localized solutions are also constructed by directly solving full field
equations \cite{kraus,park,aganagic,harvey} and as a result  they are not always stable. Thus even with explicit solutions the number of the
preserved supersymmetries remains sometimes unclear.

In this note, we shall classify  all the possible BPS equations in the higher dimensional Yang-Mills theories, which are eventually related to the
first order linear equations  above. In particular, we consider the super Yang-Mills theories on Euclidean space of even   dimension, $D$, which may
be obtained by a dimensional reduction of the $D=10$  $N=1$ super  Yang-Mills theory. The $D=2, 4, 6, 8$  theories are  realized as the field
theoretic description of  D2, D4, D6, and D8 branes. For the supersymmetry counting in the intersecting brane picture see
\cite{OhtaTownsend,Chen:1999bf}.

The methods we employ here is  fairly straightforward. The unbroken supersymmetries are basically determined by setting the gaugino variation of the
supersymmetry to vanish. Namely, for the unbroken supersymmetries, the gaugino transformation  should have non-vanishing kernel or the space of zero
eigenspinors. The vacuum solution, for example, is invariant under all supersymmetry transformations and so the dimension of the kernel is maximal.
We shall classify all the possible kernels  and  obtain the corresponding set of  BPS equations. This is done by  analyzing the projection operator
to the kernel and in turn the complete characterization of the above constant four form tensor naturally emerges.  One of the technical key
observations will be  to find out  the `canonical form' for the four form tensor which  enables us to figure out the corresponding projection
operator for each kernel. In six dimensions, as the four form is dual to a two form, the analysis is relatively  simple.  In eight dimensions the
general constant four form tensor has 70 independent components. We decompose them into chiral and anti-chiral sectors of   35 components. We
utilize the $\mbox{SO}(8)$ triality among ${\bf 8}_{{\rm v}}$, ${\bf 8}_{+}$, ${\bf 8}_{-}$ to show that the $\mbox{SO}(8)$ rotation which has 28
independent parameters can reduce any  sector down to 7 independent components but not simultaneously.   In general we shall prove that the generic
projection operators are built up  by the elementary building blocks which lead to the BPS equations for the minimally supersymmetric BPS states.
The higher supersymmetric BPS equations  then follow from imposing   multi-sets of the minimal  BPS equations. Our analysis holds for both
commutative and non-commutative spaces as well as for both Abelian and non-Abelian gauge groups.

In the field theory on four dimensional space the BPS configurations should satisfy self-dual or anti-self-dual equations carrying 1/2 of the
original supersymmetry.  In the six dimensional case, the minimal  supersymmetric one preserves non-chiral $1/4$ of the original supersymmetry. This
is the only genuinely six dimensional one. The 1/2 supersymmetric configurations also exist but they are four dimensional in its character. The
dimension eight shows more variety. The minimal supersymmetry is chiral 1/16. The bit of octonions appears here.  From the minimal we build up to
chiral 6/16. For the chirally mixed supersymmetries, the story is more complicated and our analysis may not be complete.

The plan of the paper is as follows: In section \ref{BPScon} we study the general property of the projection operators in even dimensions. We briefly
note the two and four dimensional cases.  In sections \ref{SIX} and \ref{EIGHT} the complete analysis on  six and eight dimensional spaces are
presented separately.   In section \ref{ENERGY}, by utilizing the results on the projection operators and the BPS equations, we present some
identities which spell the Yang-Mills Lagrangian as a positive definite term plus a total derivative term. This shows the ``energy bound" of the BPS
configurations as the positive definite term vanishes when the BPS equations are satisfied. In the section \ref{EXAMPLE} we apply our results to
some known explicit BPS configurations of finite or infinite energy on commutative or non-commutative spaces. We conclude with some remarks in
section \ref{CONCLUSION}.

There is one caution in the interpretation of the BPS equations; the fact that a certain configuration  satisfies the BPS equations of a given
fraction implies that the solution preserves {\it at least} the fraction of supersymmetries.
\newpage


\section{The BPS Condition\label{BPScon}}
The BPS state refers a field configuration which is invariant under some supersymmetry. In super Yang-Mills theories on Euclidean space of even
dimension $D$, a bosonic configuration is BPS if there exits a nonzero constant spinor, $\varepsilon$ of dimension $2^{D/2}$ such that the
infinitesimal supersymmetric transformation of the gaugino field vanishes
\begin{equation}
\delta\lambda=F_{ab}\gamma_{ab}\varepsilon =0\,. \label{bps}
\end{equation}
Such zero eigenspinors of the $2^{D/2}\times 2^{D/2}$  matrix $F_{ab}\gamma_{ab}$ form the kernel space $V$. We consider  global supersymmetry and so
we take the kernel $V$ to be independent of the spatial coordinates. The kernel $V$ could be different from the ``local" kernel of
$F_{ab}(x)\gamma_{ab}$ at each point $x$. The dimension of the kernel counts the unbroken supersymmetry of a given BPS configuration. For the vacuum
$F_{ab}=0$, the unbroken supersymmetry is maximal, while  for the non-BPS configuration $V$ is simply null.

The BPS field strength  should satisfy certain consistency conditions or the BPS equations in order to have a given number of unbroken
supersymmetries. The key tool we employ here is the projection operator $\Omega$ to the kernel space. With an orthonormal basis for the kernel,
$V=\{|l\rangle\}$, $1\leq l\leq N=\mbox{dim}\,V\leq 2^{D/2}$, the projection operator is
\begin{equation}
\Omega =\displaystyle{\sum_{l=1}^{N}}\,|l\rangle\langle l|\,,
\end{equation}
satisfying
\begin{equation}
\begin{array}{ccc}
F_{ab}\gamma_{ab}\Omega=0\,,~~&~~\Omega^{2}=\Omega\,,~~&~~\Omega^{\dagger}=\Omega\,.
\end{array}
\label{Omega1}
\end{equation}
Note that $\Omega$ is basis independent or unique, as it is essentially the identity on the kernel space.

In the  even dimensional Euclidean space the gamma matrices can be chosen to be Hermitian, $\gamma_a^{\dagger}=\gamma_a$, and the charge conjugation
matrix, $C$  satisfies
\begin{equation}
\begin{array}{lll}
\gamma_a^{T}=(\gamma_{a})^{\ast}=C^{\dagger}\gamma_{a}C\,,~~~& CC^{\dagger}=1\,. \label{charge}
\end{array}
\end{equation}
It follows from counting the number of symmetric $2^{D/2}\times 2^{D/2}$ matrices\,\cite{Scherkkugotownsend}
\begin{equation}
C^{T}=(-1)^{\frac{1}{8}D(D-2)}\,C\,.
\end{equation}
We let
\begin{equation}
\begin{array}{lll}
\gamma_{D+1}=i^{D/2}\,\gamma_{12\cdots D}\,,~&~\gamma_{D+1}{}^{\dagger}=\gamma_{D+1}\,,~&~\gamma_{D+1}^{\,2}=1\,.
\end{array}
\end{equation}
Then from
\begin{equation} \begin{array}{ll}
[\gamma_{ab},\gamma_{D+1}]=0\,,~~&~~ \gamma_{ab}C=C(\gamma_{ab})^{\ast}\,, \end{array}
\end{equation}
$|l\rangle\in V$ implies
\begin{equation}
\begin{array}{ll}
\gamma_{D+1}|l\rangle\in V\,,~~~&~~~C|l\rangle^{\ast}\in V\,,
\end{array}
\end{equation}
so that each of $\{\gamma_{D+1}|l\rangle \}$ and $\{C|l\rangle^{\ast}\}$ also  forms  an  orthonormal basis for $V$ separately. Consequently
\begin{equation}
\begin{array}{cc}
[\gamma_{D+1},\Omega]=0\,,~~~~&~~~~C\Omega^{\ast}C^{\dagger}=\Omega\,.
\end{array}
\label{Omega2}
\end{equation}
As the anti-symmetric products of the gamma matrices form a basis for the $2^{D/2}\times 2^{D/2}$ matrices, one can expand $\Omega$ in terms of
them. The first equation in (\ref{Omega2}) indicates only the even products contribute, and with $\Omega$ being Hermitian the second equation implies
that $\Omega$ should be a sum of foursome products of the gamma matrices with real coefficients
\begin{equation}
\Omega=\const\left(1+ \sum_{1\leq n\leq
\frac{D}{4}}\,\frac{1}{(4n)!}\,T_{a_{1}a_{2}\cdots a_{4n}}\gamma_{a_{1}a_{2}\cdots
a_{4n}}\right)\, .
\end{equation}
Furthermore with  the chiral and anti-chiral projection operators
\begin{equation}
P_{\pm}=\frac{1}{2}(1\pm\gamma_{D+1})\,,
\end{equation}
one can decompose the projection operator as
\begin{equation}
\begin{array}{cc}
\Omega=\Omega_{+}+\Omega_{-}\,,~~&~~\Omega_{\pm}\equiv\Omega P_{\pm}=P_{\pm}\Omega\,,
\end{array}
\label{decom}
\end{equation}
satisfying
\begin{equation}
\begin{array}{ll}
\Omega_{\pm}=\Omega_{\pm}^{\,2}=\Omega_{\pm}^{\,\dagger}\,,~&~\Omega_{\pm}\Omega_{\mp}=0\,.
\end{array}
\end{equation}
When combined with the charge conjugation,
\begin{equation}
C\Omega_{\pm}^{~\ast}C^{\dagger}=\left\{\begin{array}{ll} \Omega_{\pm}&~~\mbox{for~}D\equiv 0~\mbox{mod~}4\\
{}&{}\\ \Omega_{\mp}&~~\mbox{for~}D\equiv 2~\mbox{mod~}4\,.
\end{array}\right.
\label{D24}
\end{equation}
Essentially the action $|v\rangle\rightarrow C|v\rangle^{\ast}$ preserves the chirality in $D\equiv 0$ mod $4$ while it flips in $D\equiv 2$ mod $4$.
This implies that in $D\equiv 2$ mod $4$, the dimension $N$ of the kernel $V$ must be even and the projection operator can be written as
\begin{equation}
\begin{array}{ccc}
\Omega=\displaystyle{\sum_{l=1}^{N/2}}\, \big(\, |l_{+}\rangle\langle l_{+}|+|l_{-}\rangle\langle l_{-}|\, \big)
\,,~~&~\gamma_{D+1}|l_{\pm}\rangle=\pm|l_{\pm}\rangle\,,~~&~|l_{-}\rangle=C|l_{+}\rangle^{\ast}\,.
\end{array}
\label{D=2}
\end{equation}

Since the eigenvalues of $\Omega$ are either $0$ or $1$ and the non-trivial products of the gamma matrices are traceless, we have
\begin{equation}
N=\tr\,\Omega=\const\times 2^{D/2}\,.
\end{equation}
The constant $\const$ denotes the fraction of the unbroken supersymmetry so that $0\leq\nu\leq 1$. The $\const=0$ or $1$ cases are trivial, either
meaning the non-BPS state or the vacuum, $F_{ab}=0$.

The remaining constraint to ensure $\Omega$ be the projection operator is $\Omega^{2}=\Omega$. In the below we will focus on obtaining its general
solutions in each dimension. Once we solve the constraint completely we are able to obtain all the possible  BPS equations. With a given projection
operator, the formula,  $F_{ab}\gamma_{ab}\Omega =0$ can be expanded as a sum of the anti-symmetric products of even number of gamma matrices. The
BPS equations stem from requiring each coefficient of these products to vanish. After a brief review on $D=2,4$ cases, we explore the $D=6,8$
separately. Our analysis holds for both commutative and non-commutative spaces as well as for both Abelian and non-Abelian gauge groups.

In the $D=2$ case the projection matrix is trivial. Either $\Omega=0$ or $1$. Thus there is no BPS configuration. Of course this is well known. In
the $D=4$ case apart from the trivial ones $\Omega$ is given by the chiral or anti-chiral projection operator itself
\begin{equation}
\Omega=\textstyle{\frac{1}{2}}(1\pm\gamma_{1234})\,,
\end{equation}
which gives the usual self-dual equations
\begin{equation}
F_{ab}=\pm\textstyle{\frac{1}{2}}\,\epsilon_{abcd}F_{cd}\,.
\end{equation}

\section{On Six Dimensional Space\label{SIX}}
In $D=6$ case as the four form is dual to a two form, we can rewrite the  projection operator as
\begin{equation}
\Omega=\const\left(1+\frac{1}{2}T_{ab}\gamma_{ab}\gamma_{12\cdots6}\right)\,,
\end{equation}
from which we get
\begin{equation}
\Omega^{2}=\const^{2}\left[1+\textstyle{\frac{1}{2}}T_{ab}T_{ab}+(T_{ab}+\textstyle{\frac{1}{8}}\,
\epsilon_{abcdef}T_{cd}T_{ef})\gamma_{ab}\gamma_{12\cdots6}\right]\,.
\end{equation}
The condition  $\Omega^{2}=\Omega$ implies
\begin{equation}
\begin{array}{cc}
\const^{-1}-1=\frac{1}{2}T_{ab}T_{ab}\,,
~&~~(2-\const^{-1})T_{ab}+\frac{1}{4}\,\epsilon_{abcdef}T_{cd}T_{ef}=0\,.
\end{array}
\label{P6}
\end{equation}
For a nontrivial $\const\neq 0$, the supersymmetric condition becomes
\begin{equation}
0=F_{ab}\gamma_{ab}\Omega=\const (F_{ab}+\frac{1}{4}\epsilon_{abcdef}T_{cd}F_{ef})\gamma_{ab}- \const
(2F_{ac}T_{bc}\gamma_{ab}+F_{ab}T_{ab})\gamma_{12\cdots 6}\,,
\end{equation}
so that the BPS states satisfy
\begin{eqnarray}
&&F_{ab}+\frac{1}{4}\epsilon_{abcdef}T_{cd}F_{ef}=0\,,\label{6BPS1}\\ {}\nonumber\\
&&F_{ac}T_{bc}-F_{bc}T_{ac}=0\,,\label{6BPS2}\\ {}\nonumber\\ &&F_{ab}T_{ab}=0\,.\label{6BPS3}
\end{eqnarray}
The first equation guarantees the on shell condition, $D_{a}F_{ab}=0$ due to the Jacobi identity.    The others follow from
the first equation and the conditions (\ref{P6}) as we shall see below. \\

To solve Eq.(\ref{P6}) we  use the global $\mbox{SO}(6)$ transformations to rotate the real two form $T_{ab}$ to the block diagonal canonical form
so that the only non-vanishing components are $T_{12}=-T_{21}$, $T_{34}=-T_{43}$ and $T_{56}=-T_{65}$. In terms of these variables the constraints
(\ref{P6}) become
\begin{equation}
\begin{array}{ll}
\nu^{-1}-1=T_{12}^2 +T_{34}^2+T_{56}^2\,, ~~&~~ (\displaystyle{\frac{1}{2}}\nu^{-1} -1) T_{12}-T_{34}T_{56}=0\,,
\\{}&{}\\(\displaystyle{\frac{1}{2}}\nu^{-1} -1)  T_{34} -T_{56}T_{12}=0\,,~~&~~
(\displaystyle{\frac{1}{2}}\nu^{-1} -1)  T_{56} -T_{12}T_{34}=0\,.
\end{array}
\end{equation}
Solving the above equations is straightforward.  The nontrivial solutions  are $\nu=1/4,\, 2/4,\, 3/4$. The dimension of the kernel $N$ is then
$2,4,6$. There is no projection operator of odd dimension in six dimensions as argued after  Eq.(\ref{D24}).

For the $\const=1/4$ case the  two form tensor is
\begin{equation}
\begin{array}{ccc}
T_{12} = \alpha_1\alpha_2\,,~~&~ T_{34}=\alpha_1\,,~~&~  T_{56} = \alpha_2\,,
\end{array}
\end{equation}
where  $\alpha_1,\,\alpha_2$ are two independent signs
\begin{equation}
\alpha_1^2=\alpha_2^2=1\,.
\end{equation}
There are four possibilities  of $\alpha=(\alpha_1,\alpha_2)$  as $(++)$, $(+-)$, $(-+)$, $(--)$, and for each $\alpha$ there exists a corresponding
projection operator which is orthogonal to each other
\begin{equation}
\Omega_{\alpha}\Omega_{\alpha^{\prime}}=\delta_{\alpha\alpha^{\prime}}\Omega_{\alpha}\,.
\end{equation}
They are also complete as
\begin{equation}
\displaystyle{\sum_{\alpha}}\,\Omega_{\alpha}=1_{8\times 8}\,.
\end{equation}
Explicitly the BPS equations (\ref{6BPS1}) become
\begin{equation}
\begin{array}{ll}
\multicolumn{2}{c}{F_{12}+\alpha_{2}F_{34}+\alpha_{1}F_{56}=0\,,}\\ ~{}&{}\\
F_{13}+\alpha_{2}F_{42}=0\,,~~&~~F_{14}+\alpha_{2}F_{23}=0\,,\\ {}&{}\\
F_{15}+\alpha_{1}F_{62}=0\,,~~&~~F_{16}+\alpha_{1}F_{25}=0\,,\\ {}&{}\\
F_{35}+\alpha_{1}\alpha_{2}F_{64}=0\,,~~&~~F_{36}+\alpha_{1}\alpha_{2}F_{45}=0\,.
\end{array}
\label{6BPS}
\end{equation}
The remaining equations (\ref{6BPS2}) and (\ref{6BPS3}) are equivalent to the above equations and so do not provide any additional restriction.

To analyze the general BPS states  of $\nu=1/2,\,3/4$  we first note that $\gamma_{{\scriptscriptstyle 2}s{\scriptscriptstyle
-}\!{\scriptscriptstyle 1}\,{\scriptscriptstyle 2}s}\gamma_{123456}$, $s=1,2,3$ can be written as a linear combination of the $\nu=1/4$ projection
operators
\begin{equation}
\gamma_{{\scriptscriptstyle 2}s{\scriptscriptstyle -}\!{\scriptscriptstyle 1}\,{\scriptscriptstyle 2}s}\gamma_{123456}=\Omega_{{\scriptscriptstyle ++}}
+(-1)^{s}\Omega_{{\scriptscriptstyle +-}}+(-1)^{\frac{1}{2}s(s+1)}\Omega_{{\scriptscriptstyle -+}}+(-1)^{\frac{1}{2}s(s-1)}\Omega_{{\scriptscriptstyle --}}\,.
\end{equation}
Thus the general  $\const=N/8$  projection operator   in the  canonical form is also a linear combination, in fact  $N/2$ sum of the
$\Omega_{\alpha}$'s as in (\ref{D=2}).

The $\const=1/2$ projection operator is a sum of any two different $1/4$ BPS projection operators,
$\Omega_{1/2}=\Omega_{\alpha}+\Omega_{\alpha^{\prime}}$. The BPS equations for this $\Omega_{\frac{1}{2}}$ are naturally obtained by imposing two
sets of the $\nu=1/4$ conditions (\ref{6BPS}) with $\alpha$ and $\alpha'$. For example with $\alpha=(++)$ and $\alpha^{\prime}=(-+)$, the
non-vanishing two form components are $T_{56}=-T_{65} = 1$ so that the projection operator is $\Omega_{1/2}=\frac{1}{2}(1- \gamma_{1234})$ and the
BPS equations are those  of the four dimensional $1/2$ BPS equations on the first four indices while the rest of the field strength vanishes
\begin{equation}
\begin{array}{cccc}
F_{12}+ F_{34}=0\,,~&F_{13}+ F_{42}=0\,,~&F_{14}+ F_{23}=0\,,~&F_{a5}=F_{a6}=0\,.
\end{array}
\end{equation}
One can see a different choice of $\alpha$ and $\alpha^{\prime}$ leads to a self-dual or anti-self-dual BPS equations on a different four
dimensional subspace. Essentially   different   choices are related by  the $\mbox{O}(6)$ rotations.

The $\const=3/4$ projection operator can be built from a sum of any three different $1/4$ BPS projection operators. Each of them imposes the $1/4$
BPS equations on the field strength. Nevertheless  three sets of BPS equations are too much and  the only possible solution is the vacuum,
$F_{ab}=0$.

Attempts to construct  ADHM-like  solutions  of the  1/4 BPS equations (\ref{6BPS})  would end up with the 1/2 BPS solutions which are essentially
four dimensional in its character \cite{JHPprivate}. This is of course consistent with the string theory prediction that there exists no D0-D6 bound
state without the background $B$ field.

\section{On Eight Dimensional Space\label{EIGHT}}
In $D=8$, the  projection operator can be split to  chiral, $+$ and anti-chiral, $-$ parts
\begin{equation}
\Omega_{\pm}=\const(2P_{\pm}-\frac{1}{4!} T_{\pm abcd}\gamma_{abcd})\,,
\label{8p}
\end{equation}
where  $T_{\pm abcd}$ is a self-dual or anti-self-dual  four form tensor depending on the chirality, $\pm$
\begin{equation}
T_{\pm abcd}=\pm\frac{1}{4!}\epsilon_{abcdefgh}T_{\pm efgh}\,,
\end{equation}
so that $T_{\pm abcd}\gamma_{abcd}\gamma_{9}=\pm T_{\pm abcd}\gamma_{abcd}$.  The possible $\const$ values are $\const=N/16,~\,N=1,2,\cdots 8$.

The square of $\Omega_{\pm}$ can be simplified by using the self-duality of the  four form tensor and noticing, for example
$T_{abcd}T_{abce}\gamma_{de}=0$ due to the symmetries of the term.  We get
\begin{equation}
\Omega_{\pm}^{\,2}=\const^{2}\left[2(2+\frac{1}{4!}T_{\pm abcd}T_{\pm abcd})P_{\pm} -(\frac{1}{6}T_{\pm
abcd}+\frac{1}{8}T_{\pm abef}T_{\pm cdef})\gamma_{abcd}\right]\,.
\end{equation}
The condition $\Omega_{\pm}^{\,2}=\Omega_{\pm}$ implies
\begin{equation}
\begin{array}{l}
~~~~~~~~~~~~~~~\,\const^{-1}=2+\frac{1}{4!}T_{\pm abcd}T_{\pm abcd}\,,\\{}\\ (\const^{-1}-4)T_{\pm abcd} =T_{\pm abef}T_{\pm cdef}+T_{\pm
acef}T_{\pm dbef}+T_{\pm adef}T_{\pm bcef}\,.
\end{array}
\label{E11}
\end{equation}
On the other hand the supersymmetric condition reads
\begin{equation}
0=F_{ab}\gamma_{ab}\Omega_{\pm}=2\const (F_{ab}+\frac{1}{2}T_{\pm abcd}F_{cd})\gamma_{ab}P_{\pm}-\frac{1}{3}\const
F_{ae}T_{\pm ebcd}\gamma_{abcd}\,,
\end{equation}
so that the BPS states satisfy
\begin{eqnarray}
&&F_{ab}+\frac{1}{2}T_{\pm abcd}F_{cd}=0\,,\label{pre8BPS}\\ {}\nonumber\\
&&F_{a e}T_{\pm ebcd}+F_{be}T_{\pm ecad}+F_{ce}T_{\pm eabd}+F_{de}T_{\pm ecba}=0\,.
\end{eqnarray}
As shown in Eq.(\ref{QQ}),  the second equation follows from the first one  with the properties of the four form tensor (\ref{E11}).

Solving Eq.(\ref{E11}) utilizes    the $\mbox{SO}(8)$ triality among   ${\bf 8}_{{\rm v}}$, ${\bf 8}_{+}$, ${\bf 8}_{-}$ which will enable us to
rotate the self-dual or anti-self-dual four form tensor to a `canonical form'.  Eight dimensional Euclidean space admits Majorana spinors so that one
can take the  gamma matrices to be real and symmetric i.e. $C=1$. Further, the choice, $\gamma_{9}={\big(\!{\tiny\begin{array}{l}
1\,\,\,\,\,0\\0-\!\!1\end{array}}\!\big)}$ forces the gamma matrices to be off block diagonal
\begin{equation}
\gamma_{a}=\left(\begin{array}{cc}0&\rho_{a}\\\rho_{a}^{T}&0\end{array}\right)\,.
\end{equation}
The real $8\times 8$ matrix $\rho_{a}$ satisfies
\begin{equation}
\rho_{a}\rho_{b}^{T}+\rho_{b}\rho_{a}^{T}=2\delta_{ab}\,,
\end{equation}
which in turn implies $\rho_{a}\!\in\!\mbox{O}(8)$ and  $\rho_{a}^{T}\rho_{b}+\rho_{b}^{T}\rho_{a}=2\delta_{ab}$. With the above choice of gamma
matrices the spinors are real, and  the traceless chiral matrix $T_{\pm}\equiv\frac{1}{4!}T_{\pm abcd}\gamma_{abcd}$  is   also real and symmetric.
Consequently $T_{\pm}$ and  $\Omega_{\pm}$  are diagonalizable by the $\mbox{SO}(8)$ transformations of the eight dimensional chiral or anti-chiral
real spinors, ${\bf 8}_{\pm}$.

The $\mbox{SO}(8)$ triality  is apparent when we  write  the ${\bf 8}_{{\rm v}}$ generators in the basis
\begin{equation}
\gamma_{ab}=\left(\begin{array}{cc}\rho_{[a}\rho_{b]}^{T}&0\\ 0&\rho_{[a}^{T}\rho_{b]}\end{array}\right)\,.
\label{triality}
\end{equation}
Clearly $\rho_{[a}\rho_{b]}^{T}$ and  $\rho_{[a}^{T}\rho_{b]}$ are the  ${\bf 8}_{+}$ and ${\bf 8}_{-}$ generators respectively.  Thus $T_{+}$ or
$T_{-}$  can be  diagonalized, though not simultaneously,  by a $\mbox{SO}(8)$ transformation  of the vectors, ${\bf  8}_{{\rm v}}$.

To proceed further it is convenient to define the following seven quantities
\begin{equation}
\begin{array}{cccc}
E_{\pm 1}=\gamma_{8127}P_{\pm}\,,~&E_{\pm 2}=\gamma_{8163}P_{\pm}\,,~&E_{\pm 3}=\gamma_{8246}P_{\pm}\,,~& E_{\pm
4}=\gamma_{8347}P_{\pm}\,,\\ {}&{}&{}&{}\\ E_{\pm 5}=\gamma_{8567}P_{\pm}\,,~&E_{\pm 6}=\gamma_{8253}P_{\pm}\,,~&
E_{\pm 7}=\gamma_{8154}P_{\pm}\,.&{}
\end{array}
\label{Edef}
\end{equation}
Here we organize the  subscript spatial indices  of the gamma matrices such that the three  indices after the common $8$ are  identical to those of
the totally anti-symmetric octonion structure constants (e.g. \cite{Baez})
\begin{equation}
\begin{array}{c}
~~e_{i}e_{j}=-\delta_{ij}+c_{ijk}\,e_{k}\,,~~~~~~~{i,j,k=1,2,\cdots,7}\\ {}\\
1=c_{127}=c_{163}=c_{246}=c_{347}=c_{567}=c_{253}=c_{154}~~~\mbox{(others zero)}\,.
\end{array}
\end{equation}
It is easy  to see that  $E_{\pm i}$ forms a representation for the ``square'' of the octonions
\begin{equation}
\begin{array}{cc}
E_{\pm i}E_{\pm j}=\delta_{ij}\pm c^{\,2}_{ijk}\,E_{\pm k}\,,~~&~~
E_{\pm i}\equiv\pm e_{i}\otimes e_{i}\,.
\end{array}
\label{square}
\end{equation}
As they commute each other, they  form   a maximal set of the  mutually commuting  traceless  symmetric and real  matrices of a definite chirality,
$\gamma_{9}E_{\pm i}=\pm E_{\pm i}$. Thus, again using the $\mbox{SO}(8)$ transformations one can choose  an orthonormal real basis where $E_{\pm
i}$'s are simultaneously diagonal.

All together, ${\bf 8}_{\pm}$ can transform $T_{\pm}$ to a linear combination of $E_{\pm i}$'s.  The $\mbox{SO}(8)$ triality~(\ref{triality}) then
implies that ${\bf 8}_{{\rm v}}$ can rotate the self-dual or anti-self-dual four form tensor to a `canonical form' where the non-vanishing
components are only seven as $T_{\pm 1278}$, $T_{\pm 1638}$, $T_{\pm 2468}$, $T_{\pm 3478}$, $T_{\pm 5678}$, $T_{\pm 2538}$, $T_{\pm 1548}$ up to
permutations and the duality.  This is consistent with the parameter counting as $35=28+7$, where $35$ is the number of independent self-dual or
anti-self-dual components $T_{\pm abcd}$\,, and $28$ is the dimension of $\mbox{so}(8)$.

Apparently for a general $\nu=N/16$ BPS states of a definite chirality, the corresponding projection operator $\Omega_{\pm}$ is invariant under the
$\mbox{SO}(N)\times\mbox{SO}(8-N)$ subgroup of ${\bf 8}_{\pm}$. The $\mbox{SO}(8)$ triality then shows that the self-dual or
anti-self-dual four form tensor is invariant under $\mbox{SO}(N)\times\mbox{SO}(8-N)$ subgroup of ${\bf 8}_{{\rm v}}$.\\

(1) $\nu=1/16$,  $~\mbox{SO}(7)$\\
For the minimal  case $\nu=1/16$, as shown in the appendix,  the projection operator is of the general form
\begin{equation}
\Omega_{\pm}=\frac{1}{8}\bigg[P_{\pm}\,\pm\big(\salpha_{1}\salpha_{2}E_{\pm 1}+\salpha_{1}\salpha_{3}E_{\pm
2}+\salpha_{3}E_{\pm 3}+\salpha_{2}E_{\pm 4}+\salpha_{1}E_{\pm 5}+\salpha_{1}\salpha_{2}\salpha_{3}E_{\pm
6}+\salpha_{2}\salpha_{3}E_{\pm 7}\big)\bigg]\,, \label{P1/16}
\end{equation}
where  $\alpha_{1},\,\alpha_{2},\,\alpha_{3}$ are  three independent signs
\begin{equation}
1=\alpha_{1}^{2}=\alpha_{2}^{2}=\alpha_{3}^{2}\,.
\end{equation}
The corresponding  ${1/16}$ BPS states satisfy
\begin{equation}
\begin{array}{l}
F_{12}+\salpha_{1}F_{34}+\salpha_{2}F_{56}\pm\salpha_{1}\salpha_{2}F_{78}=0\,,\\{}\\
F_{13}+\salpha_{1}F_{42}+\salpha_{3}F_{57}\pm\salpha_{1}\salpha_{3}F_{86}=0\,,\\ {}\\
F_{14}+\salpha_{1}F_{23}+\salpha_{1}\salpha_{2}\salpha_{3}F_{76}\pm\salpha_{2}\salpha_{3}F_{85}=0\,,\\{}\\
F_{15}+\salpha_{2}F_{62}+\salpha_{3}F_{73}\pm\salpha_{2}\salpha_{3}F_{48}=0\,,\\{}\\
F_{16}+\salpha_{2}F_{25}+\salpha_{1}\salpha_{2}\salpha_{3}F_{47}\pm\salpha_{1}\salpha_{3}F_{38}=0\,,\\{}\\
F_{17}+\salpha_{3}F_{35}+\salpha_{1}\salpha_{2}\salpha_{3}F_{64}\pm\salpha_{1}\salpha_{2}F_{82}=0\,,\\{}\\
\pm F_{18}+\salpha_{1}\salpha_{2}F_{27}+\salpha_{1}\salpha_{3}F_{63}+\salpha_{2}\salpha_{3}F_{54}=0\,. \label{octobps}
\end{array}
\end{equation}
They are seven BPS equations for 28 components of $F_{ab}$, each of which appears once.

Especially when $\alpha_{1}=\alpha_{2}=\alpha_{3}=1$,
\begin{equation}
\begin{array}{ccc}
T_{\pm ijk8}=\pm\,c_{ijk}~~~&~\mbox{and}~&~~~F_{i8}\pm\frac{1}{2}c_{ijk}\,F_{jk}=0\,.
\end{array}
\label{+++oct}
\end{equation}

Three independent signs leads to eight possible combinations, covering all chiral or anti-chiral spinor  spaces. For each set of
$\alpha=(\alpha_{1},\alpha_{2},\alpha_{3})$ there exists a corresponding zero eigenspinor, say $|\alpha_{\pm}\rangle$ which forms an orthonormal real
basis for the chiral or anti-chiral spinor spaces. Accordingly  the projection operator (\ref{P1/16})  can be rewritten as
$\Omega_{\pm\alpha}=|\alpha_{\pm}\rangle\langle\alpha_{\pm}|$ satisfying the orthogonal completeness
\begin{equation}
\begin{array}{cc}
\displaystyle{\sum_{\alpha}}\,\Omega_{\pm\alpha}=P_{\pm}\,,~&~
\Omega_{\pm\alpha}\Omega_{\pm\alpha^{\prime}}=\delta_{\alpha\alpha^{\prime}}\Omega_{\pm\alpha}\,.
\end{array}
\end{equation}
From Eq.(\ref{P1/16})  $E_{\pm i}$ can be expressed as a linear combination of $\Omega_{\pm\alpha}$'s and hence they are  diagonal in the basis.
Consequently the general $\nu={N/16}$ projection operator in the canonical form is an $N$ sum of the $\Omega_{\pm\alpha}$'s.  Furthermore, from the
triality,  the $\frac{8!}{N!(8-N)!}$ possibilities for the $N$ sum are equivalent to each other up to $\mbox{SO}(8)$. Higher supersymmetric BPS
states then satisfy $N$ copies of the ${1/16}$ BPS equations of different $\alpha$ choices.\\

(2) $\nu={2/16}$,  $~\mbox{SO}(2)\times\mbox{SO}(6)$\\
With the $\alpha$ choices as  $(+++),~(++-)$
\begin{equation}
\begin{array}{lll}
\multicolumn{3}{c}{F_{12}+F_{34}+F_{56}\pm F_{78}=0\,,}\\
{}&{}&{}\\
F_{13}+F_{42}=0\,,~~&~~F_{57}\pm F_{86}=0\,,~~&~~F_{15}+F_{62}=0\,,\\
{}&{}&{}\\
F_{14}+F_{23}=0\,,~~&~~F_{76}\pm F_{85}=0\,,~~&~~F_{16}+F_{25}=0\,,\\
{}&{}&{}\\
F_{73}\pm F_{48}=0\,,~~&~~F_{17}\pm F_{82}=0\,,~~&~~F_{35}+F_{64}=0\,,\\
{}&{}&{}\\
F_{47}\pm F_{38}=0\,,~~&~~F_{18}\pm F_{27}=0\,,~~&~~F_{63}+F_{54}=0\,.
\end{array}
\label{2/16}
\end{equation}

(3) $\nu={3/16}$,  $~\mbox{SO}(3)\times\mbox{SO}(5)$\\
With the $\alpha$ choices as  $(+++),~(++-),~(+-+)$
\begin{equation}
\begin{array}{lll}
F_{12}+F_{34}=0\,,~~&~~F_{13}+F_{42}=0\,,~~&~~F_{14}+F_{23}=0\,,\\
{}&{}&{}\\
F_{56}\pm F_{78}=0\,,~~&~~F_{75}\pm F_{68}=0\,,~~&~~F_{67}\pm F_{58}=0\,,\\
{}&{}&{}\\
\multicolumn{3}{c}{F_{15}=F_{26}=F_{37}=\pm F_{48}\,,}\\
{}&{}&{}\\
\multicolumn{3}{c}{F_{16}=F_{52}=F_{47}=\pm F_{83}\,,}\\
{}&{}&{}\\
\multicolumn{3}{c}{F_{17}=F_{53}=F_{64}=\pm F_{28}\,,}\\
{}&{}&{}\\
\multicolumn{3}{c}{\pm F_{18}=F_{72}=F_{36}=F_{54}\,.}
\end{array}
\end{equation}
Some relevant references include \cite{Papadopoulos:1997dg,Hiraoka:2002wm}.
\newpage

(4) $\nu={4/16}$,  $~\mbox{SO}(4)\times\mbox{SO}(4)$\\
With the $\alpha$ choices as  $(+++),~(++-),~(+-+),~(+--)$
\begin{equation}
\begin{array}{lll}
F_{12}+F_{34}=0\,,~~&~~F_{13}+F_{42}=0\,,~~&~~F_{14}+F_{23}=0\,,\\
{}&{}&{}\\
F_{56}\pm F_{78}=0\,,~~&~~F_{75}\pm F_{68}=0\,,~~&~~F_{67}\pm F_{58}=0\,,\\
{}&{}&{}\\
\multicolumn{3}{c}{F_{ab}=0~~~\mbox{for}~~~{a\in\{1,2,3,4\}}~~~{b\in\{5,6,7,8\}}\,.}
\end{array}
\end{equation}

(5) $\nu={5/16}$,  $~\mbox{SO}(5)\times\mbox{SO}(3)$\\
With the $\alpha$ choices as  $(+++),~(++-),~(+-+),~(+--),~(-++)$
\begin{equation}
\begin{array}{c}
F_{12}=F_{43}=F_{65}=\pm F_{78}\,,\\
{}\\
F_{13}=F_{24}=F_{75}=\pm F_{86}\,,\\
{}\\
F_{14}=F_{32}=F_{76}=\pm F_{58}\,,\\
{}\\
F_{ab}=0~~~\mbox{for}~~~{a\in\{1,2,3,4\}}~~~{b\in\{5,6,7,8\}}\,.
\end{array}
\end{equation}

(6) $\nu={6/16}$,  $~\mbox{SO}(6)\times\mbox{SO}(2)$\\
With the $\alpha$ choices as  $(+++),~(++-),~(+-+),~(+--),~(-++),~(-+-)$
\begin{equation}
F_{12}=F_{43}=F_{65}=\pm F_{78}\,, \label{6/16}
\end{equation}
and other components are zero.\\

(7) The  ${7/16}$ BPS states do not exist, since the seven sets of ${1/16}$ BPS equations have only  the vacuum solution,
$F_{ab}=0$ which does not break any supersymmetry.\\

We have not analyzed the generic BPS states having  both chiralities, $\Omega=\Omega_{+}+\Omega_{-}$, $\Omega_{+}\neq 0$, $\Omega_{-}\neq 0$. In
general, the global $\mbox{SO}(8)$ transformations can take only one of $\Omega_{+}$, $\Omega_{-}$ to the canonical form, but not both
simultaneously.  Nevertheless,  the special case where both projection operators are in the canonical form is  manageable from our results. One can
check that the case  involves  a dimensional reduction. For example, for $\nu=\big({\frac{1}{16}}\big)_{+}+\big({\frac{1}{16}}\big)_{-}$ with
$\alpha=(+++)$ we get
\begin{equation}
\begin{array}{ll}
F_{12}+F_{34}+F_{56}=0\,,~&~~F_{13}+F_{42}+F_{57}=0\,\\
{}&{}\\
F_{14}+F_{23}+F_{76}=0\,,~&~~F_{15}+F_{62}+F_{73}=0\,,\\
{}&{}\\
F_{16}+F_{25}+F_{47}=0\,,~&~~F_{17}+F_{35}+F_{64}=0\,,\\
{}&{}\\
F_{27}+F_{63}+F_{54}=0\,,~&~~F_{a8}=0\,.
\end{array}
\end{equation}
This is essentially seven dimensional.

\section{Energy Bound\label{ENERGY}}
Pure Yang-Mills theories in the Minkowski spacetime of the dimension $D+1$, $D$ for  space, admit no local static solution having finite energy
when  $D\neq 4$, which can be seen easily  from a  scaling argument \cite{jaffe}. However this is for the commutative case and on the non-commutative
space local static configurations can exist. In this section we present some identities for $\frac{1}{4}\tr(F_{ab}^{2})$ on both six and eight
dimensional space
which  will show the energy ``bound'' of the BPS states.\\

(1) $D=6$\\
Using  the two form tensor  solution of the minimal case $\nu={1/4}$, by noticing for example $T_{ab}T_{bc}=-\delta_{ac}$, one can obtain the
following identity for the generic configurations
\begin{equation}
\frac{1}{4}\tr( F_{ab}^{2})=\frac{1}{8}\tr(F_{ab}+\frac{1}{4}\epsilon_{abcdef}T_{cd}F_{ef}+\kappa
T_{ab}T_{cd}F_{cd})^{2}-\frac{1}{16}\epsilon_{abcdef}T_{ab}\tr (F_{cd}F_{ef})\,, \label{6F2}
\end{equation}
where $\kappa=-\frac{1}{2}\pm\frac{1}{\sqrt{6}}$.  As usual, with the convention
$F_{ab}=\partial_{a}A_{b}-\partial_{b}A_{a}-i[A_{a},A_{b}]$ the last term is topological as
\begin{equation}
-\frac{1}{16}\epsilon_{abcdef}T_{ab}\tr(F_{cd}F_{ef})=-\frac{1}{4}\epsilon_{abcdef}T_{ab}\,\partial_{c}
\tr\left(A_{d}\partial_{e}A_{f}-i\frac{2}{3}A_{d}A_{e}A_{f}\right)\,.
\end{equation}
From Eq.(\ref{P6}) the vanishing of the first term of the right hand side of (\ref{6F2}) actually implies the ${1/4}$ BPS equation (\ref{6BPS1})
itself and vice versa.  As any BPS state satisfies at least one set of the $1/4$ BPS equations,  the above equation shows the
energy bound of the generic BPS states.\\

(2) $D=8$\\
With the four form tensor of the minimal case $\nu={1/16}$, a straightforward manipulation gives using (\ref{beta}) and (\ref{beta2})
\begin{equation}
\begin{array}{ll}
\displaystyle{\frac{1}{4}\tr(F_{ab}^{2})}&=\displaystyle{\frac{1}{16}\tr\left(F_{ab}+\frac{1}{2}T_{\pm abcd}F_{cd}\right)^{2}
-\frac{1}{8}T_{\pm abcd}\tr(F_{ab}F_{cd})}\\ {}&{}\\
{}&\displaystyle{=\frac{1}{16}\tr\left(F_{ab}+\frac{1}{2}T_{\pm abcd}F_{cd}\right)^{2} -\frac{1}{2}T_{\pm abcd}\,\partial_{a}
\tr\left(A_{b}\partial_{c}A_{d}-i\frac{2}{3}A_{b}A_{c}A_{d}\right)\,.}
\end{array}
\end{equation}
As in $D=6$, since any BPS state satisfies at least one set of ${1/16}$ BPS equations,  this equation shows the energy bound of the generic BPS
states.

Apart from the above quadratic topological charge,  there is  another topological charge
\begin{equation}
Q=\frac{1}{(2\pi)^{D/2}(D/2)!\,}\int\tr(\underbrace{F\wedge F\wedge\cdots\wedge F}_{D/2~~{\scriptstyle\hbox{product}}})\,\,.
\end{equation}
which does not need to be related to the above quadratic topological charge directly.


\section{Examples in $D=6$ and $D=8$\label{EXAMPLE}}

Here we comment on the known solutions of pure Yang-Mills theories on commutative or non-commutative  spaces with gauge group $\mbox{U}(1)$ or
non-Abelian. As stated before, there is no static finite energy configurations on commutative space of dimension  higher than four.

\subsection{ Octonionic Instantons}

In eight dimensional Euclidean space,  with the gauge group $SO(7)^+$ as a subgroup of $SO(8)$, an explicit solution of the octonionic BPS equation
(\ref{+++oct})  has been found \cite{fubini}. It has infinite energy, but its four-form charge over the corresponding four dimensional hyperplane is
finite. Namely  the integration  of $\tr(F_{ab}F_{cd})\,{\rm d}x_{a}\wedge{\rm d}x_{b}\wedge{\rm d}x_{c}\wedge{\rm d}x_{d}$ over
$(x_{a},x_{b},x_{c},x_{d})$  hyperplane is finite and  in fact  proportional to $T_{\pm abcd}$.  Our analysis in the previous section tells us that
it has $1/16$ supersymmetry.

\subsection{Constant Field Strength}

One can think of a constant field strength and so they are independent of space.  Let us consider the   $\mbox{U}(1)$ theory first. Strictly
speaking, in this case there exit non-linearly realized additional supersymmetries in the super Yang-Mills theories where the gaugino transforms as
$\delta \lambda=\varepsilon^{\prime}$ with the gauge fields fixed.  As a result  any constant field strength  preserves all the  supersymmetries.
Let us then consider the $\mbox{SU}(2)$ case where all the the constant fields are diagonal i.e. $F_{ab}= f_{ab}\sigma_{3}$ with $\sigma_{3}$ being
the third Pauli matrix. The additional supersymmetries do not play any role here. In this special case one may not have to go through all the
previous projection operator analysis. Using the global rotation,  $\mbox{O}(D)$, one can block diagonalize the field strength so that
$f_{ab}\gamma_{ab}=\sum_{s=1}^{D/2}\,2 f_{{\scriptscriptstyle 2}s{\scriptscriptstyle -} \!{\scriptscriptstyle 1}\,{\scriptscriptstyle
2}s}\gamma_{{\scriptscriptstyle 2}s{\scriptscriptstyle -} \!{\scriptscriptstyle 1}\,{\scriptscriptstyle 2}s}$. As the $\gamma_{{\scriptscriptstyle
2}s{\scriptscriptstyle -}\!{\scriptscriptstyle 1} \,{\scriptscriptstyle 2}s}$,  $s=1,2,\cdots,D/2$ are commuting each other and have eigenvalues
$\pm 1$, the constant configuration, $f_{{\scriptscriptstyle 2}s{\scriptscriptstyle -} \!{\scriptscriptstyle 1}\,{\scriptscriptstyle 2}s}\equiv
f_{s}$ is  BPS if and only if
\begin{equation}
\pm f_1 \pm f_2 \pm f_3 \pm \cdots\pm f_{D/2}=0\,. \label{cBPS}
\end{equation}
The number of possible  sign combinations out of  $2^{D/2}$ choices counts the number of unbroken supersymmetries. The multiplication of all the
signs determines the chirality of the corresponding zero eigenspinor.  Due to the freedom to flip the over all signs, the number of unbroken
supersymmetries is always even. This over all sign change  leaves the chirality invariant for even $D/2$, while it flips for odd $D/2$. Surely this
is consistent with the results in section \ref{BPScon}, and  Eq.(\ref{cBPS}) corresponds to (\ref{6BPS}) or (\ref{2/16}).

More explicitly, fully using the $\mbox{O}(D)$ rotation  we arrange $f_1\ge f_2 \ge f_3 \ge \cdots\ge  f_{D/2}\ge 0$.  For $D=6$ the configuration is
BPS if $f_1-f_2-f_3=0$. As there are two possible over all signs, it corresponds to the ${1/4}$ BPS state. For $D=8$  there are several
possibilities. For $f_1>f_2\ge f_3 \ge f_4 > 0$, it can only satisfy either $f_{1}-f_{2}-f_{3}+f_{4}=0$ or $f_{1}-f_{2}-f_{3}-f_{4}=0$, which has
${1/8}$ supersymmetry with the positive or negative chirality respectively. For $f_1>f_2\ge f_3 > f_4 = 0$, it satisfies both so that $\nu={1/4}$.
For $f_1=f_2>f_3=f_4>0$, it can satisfy $f_1-f_2 \pm (f_3-f_4)=0$, which has also 4 supersymmetries and hence $\nu={1/4}$ with positive chirality.
For $f_1=f_2>f_3=f_4=0$, we have $\nu={1/2}$.  Finally for $f_1=f_2=f_3=f_4>0$, it can satisfy $f_1-f_2+f_3-f_4=0$ and others so that it has six
supersymmetries of the positive chirality. This agrees with the previous ${3/8}$ BPS equations (\ref{6/16}).

For the non-Abelian gauge group, we have $F_{ab}=F^{\alpha}_{ab}\,t^{\alpha}$ and one can not block diagonalize  all of $F_{ab}^{\alpha}$'s
simultaneously by a single $\mbox{SO}(D)$ rotation in general. Nevertheless, we point out  that if the gauge group is  semi-simple of rank two or
higher,  a constant field  strength configuration in eight dimensions  can be a $1/16$ BPS state. For example, if $[t^{1},t^{2}]=0$,  one can
construct a non-Abelian constant field strength of which the non-vanishing components are  $F^{1}_{12},F^{1}_{34},F^{1}_{56},F^{1}_{78}$ and
$F^{2}_{13},F^{2}_{42},F^{2}_{57},F^{2}_{86}$ only up to the anti-symmetric property. When $F^{1}_{ab}$ and $F^{2}_{ab}$ satisfy the first and
second  equations of the $1/16$  BPS equations (\ref{octobps}) respectively  with a unique choice of $\alpha$, it is certainly a $1/16$ BPS state.

\subsection{Non-commutative Exact Solutions}

The non-commutative space is specified by the commutation relation
\begin{equation}
[x_{a},x_{b}]=i\theta_{ab}\,.
\end{equation}
Using the global $\mbox{O}(D)$ rotation one can take  the block diagonal canonical  form for $\theta$   so that the non-vanishing components are
$\theta_{{\scriptscriptstyle 2}s{\scriptscriptstyle -}\!{\scriptscriptstyle 1}\,{\scriptscriptstyle 2}s}\equiv\theta_{s}>0$, $s=1,2,\cdots,D/2$ only
up to the anti-symmetric property.  This choice clearly manifests  the  $D/2$ pairs of harmonic oscillators, $[a_{s},a^{\dagger}_{r}]=\delta_{sr}$,
$a_{s}=\frac{1}{\sqrt{2\theta_{s}}}(x_{{\scriptscriptstyle 2}s{\scriptscriptstyle -}\!{\scriptscriptstyle 1}}+ix_{{\scriptscriptstyle 2}s})$.

On  non-commutative space  any $\mbox{U}(n)$   gauge theories  are equivalent to another, in particular to a $\mbox{U}(1)$  theory provided that all
the fields are in the adjoint representation  \cite{comments}. Hence we  restrict on the $\mbox{U}(1)$ gauge group only here without loss of
generality. Writing $A_{a}=Y_{a}-\theta^{-1}_{ab}x_{b}$ gives
\begin{equation}
F_{ab}=\theta^{-1}_{ab}-i[Y_{a},Y_{b}]\,. \label{non-commF}
\end{equation}

1) Shift Operator Solitons\\
The shift operator, $S$ in the harmonic oscillator Hilbert space is almost unitary
\begin{equation}
\begin{array}{cc}
SS^{\dagger}=1\,,~~~&~~~S^{\dagger}S=1-P_{0}\,.
\end{array}
\end{equation}
where $P_{0}$ is a projection operator of finite dimension to the Hilbert space. The shift operator solution   satisfying $D_{a}F_{ab}=0$ reads
\cite{shift,witten,park,aganagic}
\begin{equation}
\begin{array}{ccc}
Y_{a}=\theta^{-1}_{ab}\,S^{\dagger}x_{b}S~~~&~\Longrightarrow~&~~~F_{ab}=\theta^{-1}_{ab}\,P_{0}\,.
\end{array}
\end{equation}
Similar to the constant solutions, this localized field configuration is supersymmetric if and only if
\begin{equation}
\displaystyle{\pm \frac{1}{\,\theta_1}\pm\frac{1}{\,\theta_2}\pm\frac{1}{\,\theta_3}\pm\cdots\pm\frac{1}{\,\,\,\theta_{{\scriptscriptstyle
D/2}}}}=0\,,
\end{equation}
and hence the counting of the supersymmetry proceeds identically as in the constant solutions \cite{witten,Fujii:2001wp}. The energy of the solution
is
\begin{equation}
E_{0}=\displaystyle{\frac{1}{2e^{2}}(2\pi)^{D/2}\left(\prod_{s=1}^{D/2}\, \theta_s\right)\left(\frac{1}{\theta_1^2}+\frac{1}{\theta_2^2} +\cdots
+\frac{1}{\,\,\,\theta_{{\scriptscriptstyle  D/2}}^2}\right) \Tr P_0\,,} \label{shiftE}
\end{equation}
and the $D$-form topological charge is
\begin{equation}
Q_{0}=\displaystyle{\left(\prod_{s=1}^{D/2}\,\theta_s\right)} \Tr (F_{12}F_{34}\cdots F_{{\scriptscriptstyle D\!-\!1\,D}})=(-1)^{D/2}\,\Tr P_{0}\,.
\label{Q0}
\end{equation}

2) Kraus and Shigemori Solutions\\
From  (\ref{non-commF})   $D=6$ $\nu=1/4$, or $D=8$ $\nu=1/8$ BPS equations (\ref{6BPS},\ref{2/16}) are equivalent, with the positive $\alpha$
choice, to
\begin{equation}
\begin{array}{ll}
\displaystyle{\sum_{s=1}^{D/2}\,[Z_{s},\bar{Z}_{s}]=\sum_{s=1}^{D/2}\,\frac{1}{\,\theta_{s}}}\,,~~~&~~~~[Z_{s},Z_{r}]=0\,,
\end{array}
\label{ncBPS}
\end{equation}
where we complexified $Y_{a}$ so that $Z_{s}=\frac{1}{\sqrt{2}}(Y_{{\scriptscriptstyle 2}s{\scriptscriptstyle -}\!{\scriptscriptstyle
1}}+iY_{{\scriptscriptstyle 2}s})$, $s,r=1,2,\cdots,D/2$.

Kraus and Shigemori found a class of finite energy configurations  satisfying $D_{a}F_{ab}=0$ in the special case where all the $\theta_{s}$'s are
equal, $\theta_{s}=\theta$. Their solutions are specified by two non-negative numbers,  $L\geq l\geq 0$ and the energy is minimized when $L=l$. In
this case  their solution reduces to
\begin{equation}
Z_{r}=\frac{1}{\sqrt{\theta}}\,Sf(N)\,a_{r}S^{\dagger}\,,
\end{equation}
where with total number operator, $N=\sum_{s}a^{\dagger}_{s}a_{s}$,
\begin{equation}
f(N)=\sqrt{1-\prod_{j=1}^{D/2}\left(\frac{l+j}{N+j}\right)}\,,
\end{equation}
and $S$ is a shift operator satisfying $1-S^{\dagger}S=P_{l}$, a projection operator to the states, $N\leq l$.  It is
straightforward to show that this is a solution of the above  non-commutative BPS equation (\ref{ncBPS}) having
$\nu=(\frac{1}{2})^{(D-2)/2}$ supersymmetry.  Its energy is
\begin{equation}
E_{{\scriptscriptstyle K\!-\!S}}=\displaystyle{\frac{1}{8e^{2}}D(D-2)(2\pi)^{D/2}\theta^{(D-4)/2}\,\Tr P_{l}\,,} \label{KSE}
\end{equation}
while the $D$-form topological charge is
\begin{equation}
Q_{{\scriptscriptstyle K\!-\!S}}=-(-1)^{D/2}\,\Tr P_{l}\,.
\end{equation}
This is exactly the ``opposite" of $Q_{0}$ in (\ref{Q0}).\\

3) Comparison\\
Let us now compare the shift operator solutions with the K-S solutions. First from  (\ref{shiftE}) and (\ref{KSE})  we get $E_{{\scriptscriptstyle
K\!-\!S}}\geq E_{0}$. The equality holds only in $D=4$, and for $D=6,8$ the K-S solitons are heavier. In six dimensions both of them can have two
supersymmetries or 1/4 of the original one. However the $\theta_{ab}$ condition  of the K-S solutions, $\theta_{1}=\theta_{2}=\theta_{3}$ forbid the
shift operator solution from being supersymmetric, which would require $1/\theta_{1}=1/\theta_{2}+1/\theta_{3}$.   In eight dimensions both can be
supersymmetric  at the same time. The shift operator solution carries $3/8$ of the original supersymmetry while the K-S solution carries $1/8$. As
their topological charges are ``opposite", they can be thought as $D0-D8$ and $\overline{D0}-D8$ systems.

\section{Conclusion\label{CONCLUSION}}

In  this note, we considered the maximally supersymmetric pure Yang-Mills theories on Euclidean space of six and eight dimensions. We determine, in
a systematic way,  all the possible fractions of the supersymmetry preserved by the BPS states and present the corresponding BPS equations. In six
dimensions, the intrinsic one has 1/4 of supersymmetry while the   1/2 BPS states here are essentially four dimensional. In eight dimensions, the
minimal fraction is chiral 1/16. Within chiral or anti-chiral sector one could build up to 6/16 without dimensional reduction. We have applied our
BPS equations to various known solutions, counted the numbers of supersymmetries and compared their topological charges.

In our eight dimensional analysis  we decomposed the constant four form tensor into chiral and anti-chiral sectors as the BPS conditions work
separately. Using the $\mbox{SO}(8)$ triality, we were able to bring any   chiral sector into the canonical form, but not both simultaneously.
Considering both chiralities in the canonical form simultaneously, we have obtained the BPS equations that are essentially seven dimensional in its
character. There is no reason not to consider the generically  mixed chiralities  where only one sector is in the canonical form. Further
clarification is necessary in this direction.

As we considered  the Euclidean pure Yang-Mills theories in even dimensions, we turned on only the magnetic field
strength for the D brane systems and  did not included the effect of electric fields. One such example involving the
electric fields  is the BPS equations describing dyons in three spatial dimensions.  In this respect it would be
interesting to generalize our results to the case where non-vanishing electric fields are allowed. Others we did not
consider includes  the super Yang-Mills theory of nine spatial dimensions.  From the view point of the dimensional
reduction, this corresponds to the eight dimensional theory plus one adjoint Higgs. Our project will be complete  in some
sense if one is able to classify all the possible  BPS equations in  the $9+1$ dimensional super Yang-Mills theory.
Nevertheless our eight dimensional results are ready for the reduction to the lower dimensions to generate some adjoint
Higgs.\\

{\large\bf Acknowledgements} We  wish to thank Piljin Yi for the valuable conversation and  Sang-Heon Yi for pointing out
the relevance of the K-S solutions to our BPS equations. This work is supported in part by KOSEF 1998 Interdisciplinary
Research Grant 98-07-02-07-01-5.

\newpage
\appendix
\begin{center}
\large{\textbf{Appendix}}
\end{center}
\setcounter{equation}{0}
\renewcommand{\theequation}{A.\arabic{equation}}
There is an identity for  an arbitrary self-dual or anti-self-dual four form tensor in $D=8$.  From the relation
\begin{equation}
\begin{array}{ll}
T_{\pm acde}T_{\pm bcde}&=\left(\displaystyle{\frac{1}{4!}}\right)^{2}\epsilon_{acdefghi}\epsilon_{bcdejklm}
T_{\pm fghi}T_{\pm jklm}\\ {}&{}\\
{}&=\displaystyle{\frac{1}{4}}\,\delta_{ab}T_{\pm cdef}T_{\pm cdef}-T_{\pm acde}T_{\pm bcde}\,,
\end{array}
\end{equation}
we obtain
\begin{equation}
T_{\pm acde}T_{\pm bcde}=\displaystyle{\frac{1}{8}}\,\delta_{ab}\,T_{\pm cdef}T_{\pm cdef}\,.
\label{Fidentity}
\end{equation}
With the four form tensor in $D=8$ projection operator  we set
\begin{equation}
Q_{abcd}=F_{ae}T_{\pm ebcd}+F_{be}T_{\pm ecad}+F_{ce}T_{\pm eabd}+F_{de}T_{\pm ecba}\,.
\end{equation}
{}From  the identity (\ref{Fidentity}) and the properties  of the four form tensor (\ref{E11}) which come from
$\Omega_{\pm}^{\,2}=\Omega_{\pm}$  we obtain
\begin{equation}
\begin{array}{ll}
0~\leq ~Q_{abcd}Q_{abcd}&=4F_{ab}F_{ac}T_{\pm bdef}T_{\pm
cdef}-12F_{ab}T_{\pm bcde}F_{cf}T_{\pm fade}\\ {}&{}\\
{}&=12(\const^{-1}-2)F_{ab}F_{ab}-6\bigg[F_{ab}F_{cd}T_{\pm
abef}T_{\pm cdef}+(4-\const^{-1}) F_{ab}F_{cd}T_{\pm abcd}\bigg]\\
{}&{}\\
{}&=12(F_{ab}+\textstyle{\frac{1}{2}}T_{\pm abcd}F_{cd})\bigg[(\const^{-1}-2)F^{ab}-T_{\pm abef}F_{ef}\bigg]\,.
\end{array}
\label{QQ}
\end{equation}
Therefore   Eq.(\ref{pre8BPS}) makes $Q_{abcd}$ vanish.\newline

Henceforth  we solve  $\Omega_{\pm}=\Omega^{\,2}_{\pm}$ for the case $D=8$, $\const={1/16}$. With the canonical choice of  the self-dual or
anti-self-dual four form tensor  we write
\begin{equation}
\Omega_{\pm}=\displaystyle{\frac{1}{8}P_{\pm}\left(1\pm\,\sum_{i=1}^{7}\,\beta_{i}E_{\pm i}\right)}\,,
\end{equation}
where
\begin{equation}
\begin{array}{lll}
\beta_{1}=T_{\pm 3456}=\pm T_{\pm 1278}\,,~&\beta_{2}=T_{\pm 2475}=\pm T_{\pm 1638}\,,~&\beta_{3}=T_{\pm 1357}=\pm T_{\pm
2468}\,,\\{}&{}&{}\\ \beta_{4}=T_{\pm 1256}=\pm T_{\pm 3478}\,,~&\beta_{5}=T_{\pm 1234}=\pm T_{\pm
5678}\,,~&\beta_{6}=T_{\pm 1476}=\pm T_{\pm 2538}\,,\\{}&{}&{}\\ \beta_{7}=T_{\pm 2376}=\pm T_{\pm 1548}\,.&{}&{}
\end{array}
\label{beta}
\end{equation}
Direct calculation using Eq.(\ref{square}) gives
\begin{equation}
\Omega_{\pm}^{\,2}=\displaystyle{\frac{1}{64}}P_{\pm}\left( 1+\sum_{i=1}^7\,\beta_{i}^{2} \pm (2\beta_{i}+
\sum_{j,k}\, c^2_{ijk}\beta_{j}\beta_{k})E_{\pm i} \right)\,.
\end{equation}
As the square of the projection operator is itself, we obtain eight equations to solve
\begin{eqnarray}
&&7-\displaystyle{\sum_{i=1}^{7}}\,\beta_{i}^{2}=0\,,\label{saturation}\\ {}\nonumber\\ &&
6\beta_{i}-\displaystyle{\sum_{j,k}\,}c^2_{ijk}\beta_{j}\beta_{k}\equiv\kappa_{i}=0\,.\label{seven}
\end{eqnarray}
Among the latter seven, with seven distinct indices $(i,j,k,l,m,r,s)$, typical four of ${\kappa_{i}\pm\kappa_{j}=0}$, ${\kappa_{m}\pm\kappa_{r}=0}$
give
\begin{equation}
\begin{array}{l}
(3\pm\beta_{k})(\beta_{i}\mp\beta_{j})=(\beta_{l}\mp\beta_{s})(\beta_{m}\mp\beta_{r})\,,\\{}\\
(3\pm\beta_{k})(\beta_{m}\mp\beta_{r})=(\beta_{l}\mp\beta_{s})(\beta_{i}\mp\beta_{j})\,,
\end{array}
\end{equation}
which in turn imply
\begin{equation}
\bigg[(3\pm\beta_{k})^{2}-(\beta_{l}\mp\beta_{s})^{2}\bigg](\beta_{i}\mp\beta_{j})=0\,.
\end{equation}
If $\beta_{i}^{2}\neq\beta_{j}^{2}$ then  $9+\beta_{k}^{2}=\beta_{l}^{2}+\beta_{s}^{2}$ and $3\beta_{k}+\beta_{l}\beta_{s}=0$ so that     either
$\beta_{l}^{2}=9$ or $\beta_{s}^{2}=9$. However, this violates the constraint on the size of $\beta_{i}^{2}$ (\ref{saturation}). Thus
$\beta_{i}^{2}=\beta_{j}^{2}$ and hence,  in general,  all the coefficients are of the same norm.  From (\ref{saturation}) we get  $\beta_{i}^{2}=1$
for all $i=1,2,\cdots,7$.  Now noticing $\beta_{i}\beta_{j}\beta_{k}=1\,~\mbox{or}~-1$,  the seven equations (\ref{seven}) reduce to
\begin{equation}
1=\beta_{1}\beta_{2}\beta_{7}=\beta_{1}\beta_{6}\beta_{3}=\beta_{1}\beta_{5}\beta_{4}
=\beta_{2}\beta_{5}\beta_{3}=\beta_{2}\beta_{4}\beta_{6}=\beta_{3}\beta_{4}\beta_{7}=
\beta_{5}\beta_{6}\beta_{7}\,.
\label{beta2}
\end{equation}
Essentially there remain three independent signs. Our choice of the solution is $\beta_{5}=\alpha_{1}$,
$\beta_{4}=\alpha_{2}$, $\beta_{3}=\alpha_{3}$ and
\begin{equation}
\begin{array}{llll}
\beta_{1}=\alpha_{1}\alpha_{2}\,,~&\beta_{2}=\alpha_{1}\alpha_{3}\,,&\beta_{6}=\alpha_{1}\alpha_{2}\alpha_{3}\,,
&\beta_{7}=\alpha_{2}\alpha_{3}\,.
\end{array}
\label{alpha}
\end{equation}

\end{document}